\documentclass[twocolumn,epjc3]{svjour3}  
\RequirePackage{graphicx}
\usepackage{siunitx}
\usepackage{microtype}
\usepackage{float}
\usepackage{amsmath,amssymb}
\usepackage{lineno}
\usepackage{multicol}
\usepackage{url}
\usepackage[colorlinks=true, linkcolor=blue, urlcolor=blue, citecolor=blue]{hyperref}
\usepackage[style=numeric-comp,backend=bibtex,maxnames=3,minnames=1,sorting=none,giveninits=true]{biblatex}

\journalname{Eur. Phys. J. C}

\begin{document}

\title{Seasonal Variations of the Atmospheric Muon Neutrino Spectrum measured with IceCube
}

\onecolumn

\author{R. Abbasi\thanksref{loyola}
\and M. Ackermann\thanksref{zeuthen}
\and J. Adams\thanksref{christchurch}
\and S. K. Agarwalla\thanksref{madisonpac,a}
\and J. A. Aguilar\thanksref{brusselslibre}
\and M. Ahlers\thanksref{copenhagen}
\and J.M. Alameddine\thanksref{dortmund}
\and N. M. Amin\thanksref{bartol}
\and K. Andeen\thanksref{marquette}
\and C. Arg{\"u}elles\thanksref{harvard}
% ...
}
\authorrunning{IceCube Collaboration}
\thankstext{a}{also at Institute of Physics, Sachivalaya Marg, Sainik School Post, Bhubaneswar 751005, India}
\thankstext{b}{also at Department of Space, Earth and Environment, Chalmers University of Technology, 412 96 Gothenburg, Sweden}
\thankstext{c}{also at Earthquake Research Institute, University of Tokyo, Bunkyo, Tokyo 113-0032, Japan}
\institute{
% ...
Universit{\'e} Libre de Bruxelles, Science Faculty CP230, B-1050 Brussels, Belgium \label{brusselslibre}
% ...
\and Department of Physics and Laboratory for Particle Physics and Cosmology, Harvard University, Cambridge, MA 02138, USA \label{harvard}
% ...
\and Department of Physics, Loyola University Chicago, Chicago, IL 60660, USA \label{loyola}
\and Dept. of Physics and Astronomy, University of Canterbury, Private Bag 4800, Christchurch, New Zealand \label{christchurch}
% ...
\and Niels Bohr Institute, University of Copenhagen, DK-2100 Copenhagen, Denmark \label{copenhagen}
\and Dept. of Physics, TU Dortmund University, D-44221 Dortmund, Germany \label{dortmund}
% ...
\and Dept. of Physics and Wisconsin IceCube Particle Astrophysics Center, University of Wisconsin{\textemdash}Madison, Madison, WI 53706, USA \label{madisonpac}
% ...
\and Department of Physics, Marquette University, Milwaukee, WI 53201, USA \label{marquette}
% ...
\and Bartol Research Institute and Dept. of Physics and Astronomy, University of Delaware, Newark, DE 19716, USA \label{bartol}
% ...
\and Deutsches Elektronen-Synchrotron DESY, Platanenallee 6, D-15738 Zeuthen, Germany \label{zeuthen}
}


\date{Received: date / Accepted: date}

\maketitle
\twocolumn
\begin{abstract}

This study presents an analysis of seasonal variations in the atmospheric muon neutrino flux, using 11.3 years of data from the IceCube Neutrino Observatory. By leveraging a novel spectral unfolding method, we explore the energy range from \SI{125}{\giga\electronvolt} to \SI{10}{\tera\electronvolt} for zenith angles from \SIrange{90}{110}{\degree}, corresponding to the Antarctic atmosphere. Our findings reveal that the differential measurement of the amplitudes of the seasonal variation is consistent with an energy-dependent decrease reaching ($-4.5$ ± 1.2)\% during Austral winter and increase to (+3.9 ± 1.3)\% during Austral summer relative to the annual average at \SI{10}{\tera\electronvolt}. While the unfolded flux exceeds the model predictions by up to 30\%, the differential measurement of the seasonal to annual average flux remains unaffected. The measured seasonal variations of the muon neutrino spectrum are consistent with theoretical predictions using the MCEq code and the NRLMSISE-00 atmospheric
model. 

\end{abstract}

\section{Introduction}
\label{intro}
\begin{figure}
    \centering
    \includegraphics[width=0.4\textwidth]{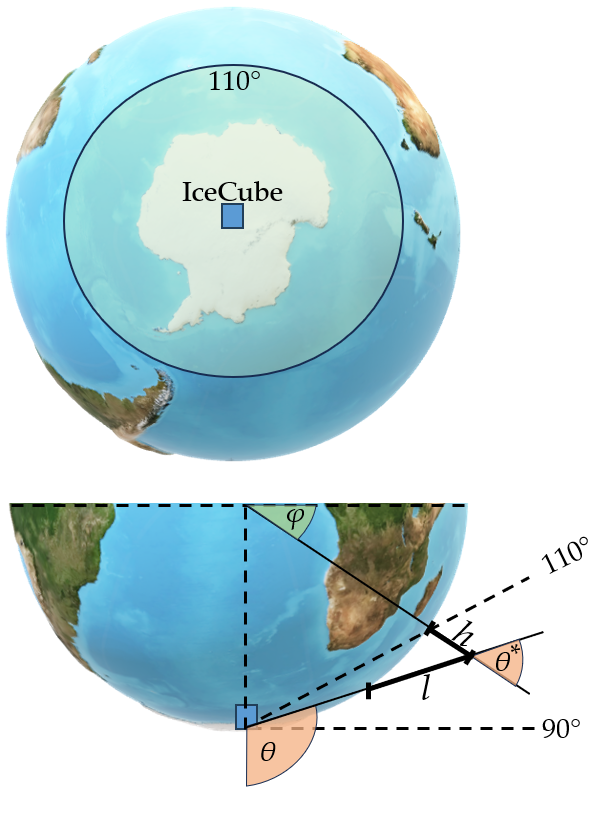}
    \caption{Illustration of the experimental setup. The upper figure depicts the atmospheric region, the Antarctic atmosphere, from which the neutrinos are analyzed in this paper, and the IceCube detector at the South Pole. The illustration at the bottom shows the definition of relevant physical quantities and the zenith region to be analyzed in IceCube coordinates $\theta$. The path of the neutrino along the line of sight is given by $l$, the vertical height of the atmosphere by $h$. The spanned zenith region in the bottom figure is not to scale.}
    \label{fig:angles}
\end{figure}

Atmospheric leptons, such as muons, muon neutrinos, and electron neutrinos, originate from highly relativistic meson decays, mainly from kaons and pions, within cosmic-ray-induced particle cascades (air showers) \cite{Gaisser:2016uoy}. If the interaction length $\lambda_\text{int}$ of a parent meson exceeds its decay length $\lambda_\text{dec}$, the particle is more likely to decay before undergoing an inelastic collision. In the case of a charged pion, it will distribute its entire energy between a muon and a muon neutrino. Conversely, if $\lambda_\text{int} \ll \lambda_\text{dec}$, a parent meson is more likely to undergo an inelastic interaction with an air nucleus. This produces lower-energy mesons and decay products, resulting in a steeper spectral index of the atmospheric lepton spectrum. Since $\lambda_\text{int} \propto 1/\rho$ holds for interaction length expressed in units of distance, this balance is sensitive to the local atmospheric density $\rho(l)$, which varies with altitude, season, and temperature. The energy at which the probabilities of meson decay and interaction are equal is referred to as the critical energy. 
This transition occurs around the critical energy of approximately \SI{115}{\giga\electronvolt}$/\cos{\theta^*}$ for charged pions and \SI{857}{\giga\electronvolt}$/\cos{\theta^*}$ for charged kaons \cite{Desiati:2010wt}, with $\theta^*$ measured with respect to the line perpendicular to the Earth's surface at the point of entry of the cosmic ray into the atmosphere, as illustrated in Fig.~\ref{fig:angles}. 

Atmospheric density and temperature are critical parameters in this process. Seasonal temperature variations in the upper stratosphere and lower thermosphere modify the atmospheric density profile, which in turn affects the competition between meson decay and interaction. In denser regions of the atmosphere, the probability of interaction increases, making it less likely for mesons to decay before undergoing an inelastic collision. This corresponds to a higher value for the critical energy, thereby suppressing the production of atmospheric muons and neutrinos at lower energies. As a result, temporal and spatial variations in atmospheric density—driven by seasonal changes or geographic location—lead to observable effects on the spectral index and the seasonal modulation of atmospheric lepton fluxes.

Variations in atmospheric conditions and their impact on air showers can be investigated using neutrinos and muons.
The measurement of seasonal variations in the muon flux is well-established for underground particle observatories \cite{Barrett:1952woo,Sagisaka:1986bq,MACRO:1997teb,Bouchta:1999kg,Adamson:2014xga,Tilav:2010hj,DoubleChooz:2016sdt,OPERA:2018jif,Borexino:2018pev}. In contrast, measuring seasonal variations in the atmospheric neutrino flux poses significant challenges due to the limited event rate and substantial background from cosmic ray muons. The IceCube detector mitigates this challenge by being shielded from muons by the Earth when observing in the northerly direction (see Fig.~\ref{fig:angles}), while neutrinos, capable of penetrating the Earth, can reach the detector from all directions. 
The effect of seasonal variations provides information on the K/$\pi$-ratio, due to the difference in the critical energy of these mesons \cite{Grashorn:2009ey,Verpoest:2024dmc}. Whereas variations in the muon flux are predominantly due to pions, variations in the  neutrino flux are primarily driven by contributions from both pions and kaons \cite{Fedynitch:2018cbl}. Determining seasonal variations in the neutrino flux can constrain kaon production in air showers, a major source of uncertainty in atmospheric neutrino flux predictions \cite{Barr:2006it,Fedynitch:2017trv}. Furthermore, these measurements probe a larger atmospheric region compared to muon-based observations in IceCube and provide critical background estimates for astrophysical neutrino searches. Using six years of experimental data from the IceCube Neutrino Observatory, the amplitude of seasonal variations in the atmospheric neutrino flux and its correlation with temperature were determined in \cite{IceCube:2023qem} as proposed in \cite{Ackermann:2005qf,Heix:2019jib}.

This study presents a measurement of the seasonal variations of the muon neutrino energy spectrum using 11.3 years of data from the IceCube Neutrino Observatory, effectively doubling the size of the data set used previously in \cite{IceCube:2023qem}. The increased event count allows us to analyze seasonal variations in the energy spectrum with a spectrum unfolding method. 

This paper is structured as follows. The energy-dependence of seasonal variations is discussed in Section \ref{sec:theory}, the IceCube Neutrino Observatory and the event selection are introduced in Section \ref{sec:data}. Section \ref{sec:ratevariation} presents the seasonal variation observed in the event rate in IceCube. The spectrum unfolding method is introduced in Section \ref{sec:unfolding} with the final results presented in Section \ref{sec:results}. 

\section{Seasonal variations of atmospheric neutrinos} \label{sec:theory}

\subsection{Theoretical modeling} \label{sec:mceq}

The Matrix-Cascade Equations (MCEq) code \cite{Fedynitch:2015zma, Fedynitch:2018cbl} is a numerical solver of the one-dimensional relativistic transport equations (cascade equations) for particles propagating through dense or gaseous media, such as the Earth's atmosphere. MCEq iteratively calculates the evolution of the individual spectra of secondary particles produced through hadronic interactions or decays as a function of the slant depth in the atmosphere, given by the path integral of the atmospheric density  along the particle trajectory:
$
X = \int_l^{\infty} \rho(l^{\prime}) \mathrm{d}l^{\prime} $.
The main required inputs are a model for the spectrum of primary cosmic rays at the top of the atmosphere, a model for the atmosphere's density, and a parameterization of the cross section for inclusive particle production and decays. At energies above \SI{10}{\giga\electronvolt}, low-energy effects such as solar modulation and deflections in the Earth's magnetic field can be safely neglected \cite{Gaisser:2002jj,Honda:2015fha}.

In this work, the initial cosmic ray flux is parameterized by the H3a \cite{Gaisser:2011klf} model, and SIBYLL-2.3c \cite{Fedynitch:2018cbl,Riehn:2019jet} is used as the hadronic interaction model for all comparisons within this paper.

We test three different atmospheric density models. The default in MCEq for the location of the South Pole is the numerical representation of the NRLMSISE-00 model \cite{msis}. It is an empirical static atmospheric model that integrates a comprehensive array of measurements, including satellite data from the Atmospheric Explorer and Dynamics Explorer and ground-based observations. It combines data from mass spectrometers, accelerometers, and other instruments to precisely represent atmospheric temperature, density, and composition from the Earth's surface to the exosphere. The calculation of atmospheric cascades in MCEq extends to heights of \SI{112.8}{\kilo\metre}.

We also tested two parameterizations from \cite{IceCube:2023qem} based on temperature measurements by the Atmospheric Infrared Sounder (AIRS) \cite{airs} and by the European Centre for Medium-Range Weather Forecasts (ECMWF) \cite{hersbach2023era5}, which includes an interpolation model of data taken by other instruments. The AIRS instrument orbits the Earth daily, resulting in $14\times 2$ measurements per day. The temperature across the different positions around Earth is determined for 24 pressure levels ranging from \SIrange{1000}{1}{\hecto\pascal}, providing an angular resolution of \SI{1}{\degree} $\times$ \SI{1}{\degree} in both latitude and longitude. Multiple gaps and missing data points exist due to the limited swath size and calibration processes in which the instrument cannot take data. These gaps are filled by interpolating the previous and consecutive measurement. The temperature profiles are converted into density, assuming the ideal gas equation. These density profiles are inputs for MCEq to obtain the daily forecast of the atmospheric neutrino rate from April 2012 to April 2017, which is then averaged into monthly rates. The ECMWF parameterization is based on the ERA-5 reanalysis dataset covering data from April 2012 to 2013. Due to limited statistical power, we cannot clearly distinguish the year-to-year variations of monthly neutrino rates. Therefore, the periods of the AIRS and ECMWF parameterizations are calculated as an average across the years, as atmospheric temperature variations are dominated by an annual cycle.

In addition, we introduce the daemonflux model \cite{Yanez:2023lsy}, which utilizes the MCEq framework, here using \newline NRLMSISE-00 as atmospheric parameterization, to calculate fluxes from calibrated muon data, employing the data-driven hadronic interaction model DDM \cite{Fedynitch:2022vty} and the Global Spline Fit cosmic-ray composition model \cite{Dembinski:2017zsh}.

\subsection{Neutrino production in the Antarctic atmosphere} \label{sec:atmosphere}

The atmospheric region investigated corresponds to horizontally arriving neutrinos at the IceCube Neutrino Observatory at the South Pole, shown in the upper part of Fig.~\ref{fig:angles}. The lower part illustrates the geometry of the arriving neutrinos at the detector. The Antarctic atmosphere is  characterized by a persistent and large-scale cyclone above continental Antarctica and parts of the Southern Ocean centered near the pole, referred to as a polar vortex, which extends into the troposphere and the stratosphere \cite{2022ACP....22.4187L}. The polar vortex strengthens in winter and weakens in summer. The stratospheric vortex is located at altitudes from \SIrange{10}{50}{\kilo\metre}, driven by the temperature gradient between the Equator and the poles. The clockwise winds confine cold air within the vortex, preventing it from mixing with the warmer surrounding air. The polar vortex remains confined until the heating of the stratosphere starts in August/September, triggered by the sunrise above Antarctica. Sudden stratospheric warmings can cause the vortex to break, splitting it into two parts, during which temperatures can rise dramatically -- by up to \SI{60}{\kelvin} in the stratosphere within a week. This effect has been observed using muons in IceCube \cite{Tilav:2010hj}.

\begin{figure}
    \centering
    \includegraphics[width=0.5\textwidth]{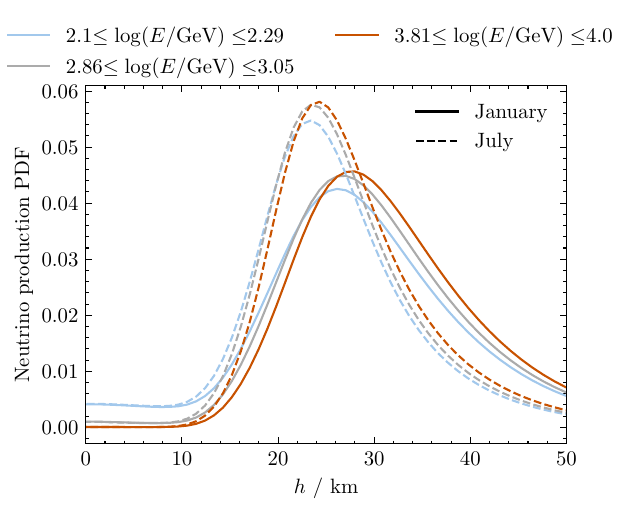}
    \caption{Neutrino production profiles for January 1st (solid) and July 1st (dashed) for the neutrino sample from the IceCube Neutrino Observatory, integrated over the zenith range from \SIrange{90}{110}{\degree} calculated using MCEq with the NRLMSISE-00 atmospheric model. The profiles in this figure are displayed for the first, fifth and highest energy bin.}
    \label{fig:prod_height}
\end{figure}

The production profiles for neutrinos that pass the event selection in the analyzed dataset, introduced in Section \ref{sec:eventselection}, are obtained by calculating neutrino spectra $\Phi_{\nu_\mu + \bar{\nu}_\mu}(E, h)$ with MCEq in steps of slant depth $X$ and altitude $h$.  The same energy binning is used as in the subsequent analysis, using ten equidistant logarithmic energy bins from \SI{125}{\giga\electronvolt} to \SI{10}{\tera\electronvolt} (see Section \ref{sec:proxies} and  Table \ref{tab:flux}). These spectra are then folded with the effective areas \cite{Stettner2021} to derive event rates $N_j$ in each energy bin, which are subsequently numerically integrated and normalized to a production profile probability density function.  The integration over the solid angle has been done by computing effective areas and neutrino fluxes and summing over five equidistant zenith bins in $\cos(\theta)$ covering the analysis angular range from $
\theta = \SIrange{90}{110}{\degree}$.

As shown in Fig.\ \ref{fig:prod_height} for the first, fifth and last energy bins within the analyzed range, the neutrino production mainly occurs at altitudes from \SIrange{15}{50}{\kilo\metre}, within the polar vortex. The mean and 68\% quantile range of neutrino production altitudes for these energy bins and seasons are displayed in Table \ref{tab:altitude}.

During Austral winter, represented by July 1st, the production profiles are narrower due to the denser atmosphere. In this season, the average neutrino production in the first energy bin ($2.1 \leq \log(E/\si{\giga\electronvolt}) \leq 2.29$) occurs at an altitude of \SI{25.4}{\kilo\metre}, with the 68\% quantile range spanning from \SI{18.0}{\kilo\metre} to \SI{32.7}{\kilo\metre}. For the highest energy bin ($3.81 \leq \log(E/\si{\giga\electronvolt}) \leq 4.0$), the mean production altitude is \SI{27.6}{\kilo\metre}, with the 68\% quantile range from \SI{20.7}{\kilo\metre} to \SI{34.6}{\kilo\metre}. Conversely, during Austral summer, the atmosphere expands, leading to higher-altitude production. For the lowest energy bin, the 68\% quantile range of neutrino production altitudes is from \SI{19.8}{\kilo\metre} to \SI{38.7}{\kilo\metre}, and for the highest energy bin, from \SI{22.5}{\kilo\metre} to \SI{40.0}{\kilo\metre}.
At altitudes below \SI{10}{\kilo\metre}, production at energies below a few \si{\tera\electronvolt} is attributed primarily to neutrinos from muon decay (see Fig.~\ref{fig:spectrum_parent}).

\begin{table*}[h]
\centering
\begin{tabular}{lcccc}
\hline
 & \multicolumn{2}{c}{January} & \multicolumn{2}{c}{July} \\ 
Energy Bin & Mean / \si{\kilo\metre} & 68\% Quantile Range / \si{\kilo\metre} & Mean / \si{\kilo\metre} & 68\% Quantile Range / \si{\kilo\metre} \\ \hline

$~2.1 \leq \log(E/\si{\giga\electronvolt}) \leq 2.29$ & 28.8 & 19.8 -- 38.7 & 25.4 & 18.0 -- 32.7 \\
$2.86 \leq \log(E/\si{\giga\electronvolt}) \leq 3.05$ & 30.2 & 21.6 -- 38.7 & 26.6 & 19.8 -- 33.6 \\
$3.81 \leq \log(E/\si{\giga\electronvolt}) \leq 4.0$ & 31.4 & 22.5 -- 40.0 & 27.6 & 20.7 -- 34.6 \\ \hline
\end{tabular}
\caption{Mean and 68\% quantile range of neutrino production altitude for January 1st and July 1st for the first, fifth, and highest energy bins, respectively.}
\label{tab:altitude}
\end{table*}

It can be noted that the neutrino production height is energy-dependent with higher-energy neutrinos being produced at higher altitudes. The seasonal variation in production heights, from summer to winter at a given energy, is stronger than the variation across different energy levels within a single season.

\subsection{Expected seasonal variations of the atmospheric neutrino rate and the spectrum}

\begin{figure}
\centering
\includegraphics[width=\columnwidth]{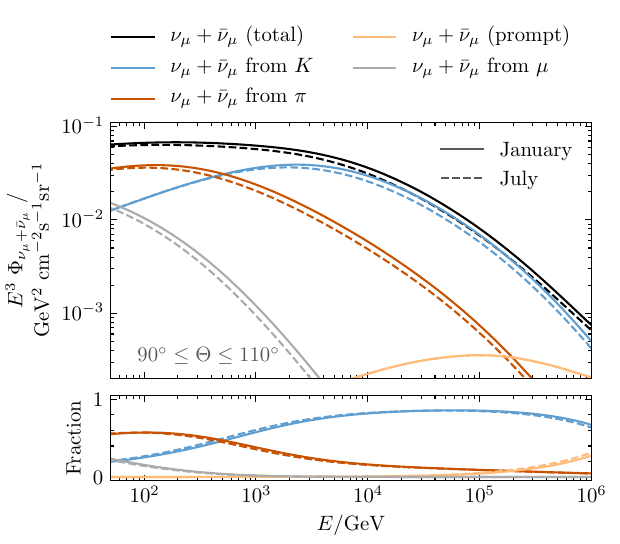}
\caption{Upper panel: Neutrino spectrum calculated with MCEq, showing contributions from parent particles for two seasons, January 1st and July 1st. The seasonal differences in the total spectrum increase with energy above the critical energy of the respective parent particle. Lower panel: The fraction of the total flux, where the prompt flux component does not show seasonal differences because of its critical energy exceeding the displayed energy range.}
\label{fig:spectrum_parent}
\end{figure}
\begin{figure}
\centering
\includegraphics[width=\columnwidth]{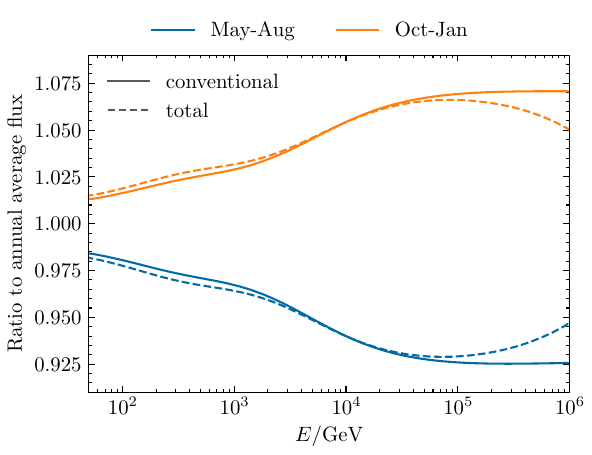}
\caption{The calculated ratio of the seasonal muon neutrino flux to the annual average, using MCEq with NRLMSISE-00 as the atmospheric density model, for the zenith range from \SIrange{90}{110}{\degree} during Austral summer and winter. The total flux, including the prompt component, is shown with dashed lines, while solid lines represent the conventional flux. The strength of seasonal variation is expected to decrease as prompt neutrinos dominate at energies above several \SI{100}{\tera\electronvolt}.}
\label{fig:mceq_prediction}
\end{figure}
\begin{figure}
\includegraphics[width=0.95\columnwidth]{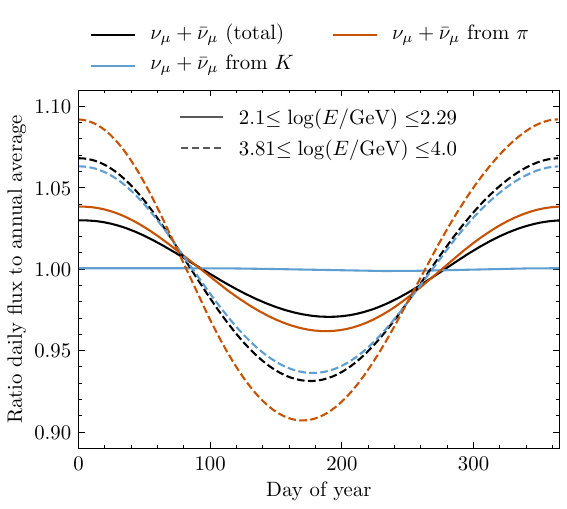}
\caption{Expected total neutrino rate variation per day of year relative to the annual average, calculated from the NRLMSISE-00 model for the zenith range from \SIrange{90}{110}{\degree}. The solid line shows the rate derived for the lowest energy bin ($2.1 \leq \log(E/\si{\giga\electronvolt}) \leq 2.29$), while dashed lines show the rate  for the highest energy bin ($3.81 \leq \log(E/\si{\giga\electronvolt}) \leq 4.0$). The rate variation is depicted for the total neutrino flux in black, while the contribution from kaons is shown in blue, and from pions in red. The seasonal variation in the first bin is dominated by pions. The contribution from kaons is compatible with the annual average as the energies are below the critical energy for kaons. The variations in the highest energy bin are mainly driven by kaons.}
\label{plot:neutrinoparent}
\end{figure}

The expected atmospheric muon neutrino flux  from different production channels in the zenith range from \SIrange{90}{110}{\degree} is displayed in Fig.~\ref{fig:spectrum_parent} for two seasons, represented by January 1st and July 1st. 
The neutrino flux from pions and kaons, known as the \textit{conventional} component, shows only small seasonal variations at energies below their critical energies. Neutrinos produced via muon decay originate from higher-energy pions, as they undergo an additional decay step compared to those produced directly from pion decay. Because the seasonal dependence of pion decay increases with energy, the resulting neutrinos from muon decay exhibit a stronger seasonal variation than neutrinos from pion decay at the same fixed energy. The \textit{prompt} neutrino flux component from short-lived mesons remains unaffected by seasonal temperature changes due to the immediate decay of their parent particles, driven by their large critical energies above $10^7$ \si{\giga\electronvolt} \cite{Desiati:2010wt}.

Above the critical energy, the spectral index becomes steeper with respect to the primary cosmic ray spectrum, as neutrino production is suppressed by the reinteraction of their parent mesons. The prompt component only contributes substantially to the total flux above several \SI{100}{\tera\electronvolt}, in particular at the horizon, flattening the total spectrum, which approximately follows the spectral index of the cosmic ray primary due to the immediate decay of the neutrino's parents.

Fig.~\ref{fig:mceq_prediction} shows the ratio of seasonal to annual average conventional and total muon neutrino fluxes for Austral summer and winter. As can be inferred from Fig.~\ref{fig:spectrum_parent}, the total flux below \SI{500}{\giga\electronvolt} is dominated by the seasonal variation of neutrinos from pion decay. Above this energy, the amplitude of variations increases at a slower rate due to the growing contribution of kaons to the total flux, which exhibit smaller seasonal variations because of their higher critical energy. The stronger rise in amplitude above the kaon critical energy is driven by the growing seasonal dependence of the kaon contribution to the neutrino flux. The amplitude approximately doubles around \SI{10}{\tera\electronvolt} over two orders of magnitude in energy. The seasonal variation of the conventional component reaches its maximum amplitude above \SI{100}{\tera\electronvolt} and remains constant beyond this. 
Below approximately \SI{4}{\tera\electronvolt}, muon decay contributes to the seasonal variation, increasing the amplitude by up to 0.5\% compared to the conventional flux variation. The total seasonal variation strength decreases above \SI{60}{\tera\electronvolt} due to the increasing contribution of seasonally independent prompt neutrinos.

The amplitude variation in the lowest and the highest energy bins within the energy range of interest from \SI{125}{\giga\electronvolt} to \SI{10}{\tera\electronvolt} is shown in Fig.~\ref{plot:neutrinoparent}. 
For the first energy bin ($2.1\leq \log(E/\si{\giga\electronvolt}) \leq 2.29$ ), the expected total amplitude is approximately $\pm 3$\%, primarily driven by variations in the pion component ($\pm 4$\%). The total variation amplitude is about 1\% lower than the amplitude attributed to seasonal variations of pions due to the contribution of neutrinos originating from muon decay at energies below \SI{1}{\tera\electronvolt}, as depicted in Fig.~\ref{fig:spectrum_parent}.
The neutrino flux from kaons remains consistent with the annual average, as the critical energy has not yet been reached.  In the highest energy bin ($3.81\leq \log(E/\si{\giga\electronvolt}) \leq 4.0$), the amplitude is dominated by kaons, whereas the smaller contribution from pions increases the amplitude from $\pm 6.5$\% (kaons) to $\pm 6.8$\% (total). The subdominant contribution of neutrinos from pion decay is expected to show a variation amplitude of more than $\pm 9$\%.

The variation phase shows a slightly slower decrease from day 0 to 180 governed by the presence of the polar vortex, followed by a slightly more rapid increase during the latter half of the year, with the lower energy bins dominated by neutrinos from pions and the higher bins by neutrinos from kaons. Differences in the timing of extrema and inflection points reflect temperature variations at different production altitudes. With sufficient statistics, seasonal variations could be more effectively used to constrain pion and kaon components in theoretical models in the future.

\section{Data}
\label{sec:data}

\subsection{The IceCube Neutrino Observatory}
\label{sec:2}

The IceCube Neutrino Observatory is a cubic-kilometer neutrino detector embedded in the glacial ice at the geographic South Pole \cite{IceCube:2016zyt}. The detection volume is instrumented by 5160 digital optical modules (DOMs) arranged on 86 cable strings on a hexagonal grid. A DOM consists of a photomultiplier tube in a pressure-resistant glass sphere and readout electronics. Neutrinos are detected indirectly via the Cherenkov light emitted by secondary particles created in interactions inside the ice or nearby bedrock. The IceCube detector is sensitive to all neutrino flavors, which can be distinguished by their signature in the detector.

\subsection{Event selection} 
\label{sec:eventselection}

The event selection follows the procedure described in \cite{IceCube:2013gge}, originally developed to obtain a high-purity and well-reconstructed sample of muon neutrino-induced tracks. This selection has been used as a standard sample for various IceCube analyses, including studies of the astrophysical neutrino flux \cite{IceCube:2021uhz} and seasonal variations of atmospheric neutrinos \cite{IceCube:2023qem}.

The event selection targets muon neutrinos interacting via charged-current interactions. The resulting muon produces Cherenkov light while traversing the detector, leading to elongated, track-like signatures. Depending on the location of the neutrino interaction, these tracks can either enter the detector from the outside (through-going) — when the neutrino interacts in the surrounding ice — or originate within the detector volume itself (starting tracks).

The main background for selecting atmospheric neutrinos consists of atmospheric muons, which exceed the detection rate of atmospheric neutrinos by six orders of magnitude. However, the Earth is a natural shield against these muons, as they can only propagate through matter for a few kilometers at the energies relevant to this analysis. Consequently, only neutrino-induced muons can reach the detector in arrival directions with zenith angles $\gtrapprox$ \SI{90}{\degree}, where the particle must enter from a horizontal direction or below, known as up-going events. The deep underground detector location provides additional shielding for zenith angles above \SI{86}{\degree}. 

A boosted decision tree (BDT) is trained to efficiently remove misreconstructed atmospheric muons that are falsely identified as up-going events. Additionally, another background arises from cascade-like events, which are characterized by a spherical Cherenkov light emission pattern. These cascades result from neutral-current interactions of all neutrino flavors or charged-current interactions of electron and tau neutrinos. To further purify the sample, a second BDT is applied, achieving a high-purity muon neutrino sample of $99.85$\%  \cite{IceCube:2016umi}.

The measurement of seasonal variations is conducted on a subset of events arriving within the zenith range from \SIrange{90}{110}{\degree}, as illustrated in Fig.\ \ref{fig:angles}. The sample, therefore, exclusively contains events from latitudes of the Southern Hemisphere with the most significant seasonal temperature changes across the globe. Larger zenith angles are excluded, as the seasonal temperature fluctuations in the stratosphere become too small towards the poles. A detailed discussion of the zenith region selection  can be found in \cite{IceCube:2023ezh}. 

The seasonal variations in the Northern Hemisphere are smaller in amplitude compared to the South, yielding smaller variations in the neutrino rate. The available statistics from the Arctic region is not yet large enough to conduct an energy-dependent seasonal analysis. Consequently, the neutrinos traversing the Earth and reaching the detector almost horizontally are particularly suitable for measuring seasonal variations in the neutrino flux due to the larger amplitude of temperature variations in the stratosphere and the characteristics of the polar vortex above Antarctica, as explained in Section \ref{sec:atmosphere}, while retaining the natural shielding against atmospheric muons \cite{IceCube:2023qem}.

No distinction can be made between neutrinos of atmospheric or astrophysical origin for each event. However, the contribution of astrophysical neutrinos is negligible at energies below \SI{10}{\tera\electronvolt} with less than $1$\% of the total neutrinos estimated to be of astrophysical origin. This analysis is conducted on 386,542 track-like neutrino-induced muons over 11.3 years of effective data-taking from May 2011 to December 2022.

\section{Observed seasonal variation of the event rate}\label{sec:ratevariation}

\begin{figure}
    \centering
    \includegraphics[width=0.49\textwidth]{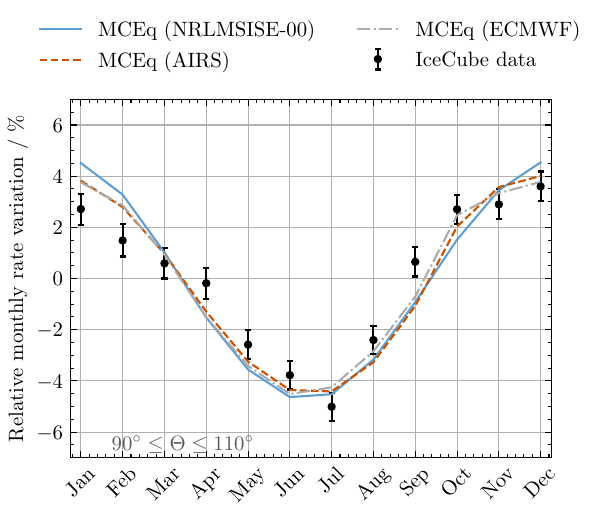}
    \caption{Relative monthly muon neutrino rate in IceCube within the zenith range from \SIrange{90}{110}{\degree}, averaged over the period from May 2011 to December 2022. Error bars represent the Poissonian uncertainty of the rate. The model expectations using three different parameterizations for the atmospheric density, described in more detail in Section \ref{sec:mceq}, are shown in different linestyles.
    }
    \label{fig:rate_variation}
\end{figure}

The observed monthly rate is determined by sorting data runs into months according to their start times within the data-taking period from May 2011 to December 2022, as explained in Section \ref{sec:eventselection}, including leap years. Subsequently, the variation of the average monthly rate is analyzed relatively to the annual average flux.

Fig.~\ref{fig:rate_variation} shows the ratio of the monthly averages over the annual average neutrino rate. The measured ratio reaches its maximum during the Austral summer in December at $(+3.6\pm 0.6)\%$, followed by a smooth decline from January to July, reaching a minimum of $(-5.0\pm 0.6)\%$ in Austral winter. The observed transition from winter to summer amplitude is more rapid than the reverse transition due to the presence of the polar vortex, which confines the cold air (discussed in Section \ref{sec:atmosphere}). This rapid transition from winter to summer contrasts with the smoother change observed in the reverse direction, with the amplitude of the winter-to-summer shift being steeper than what is predicted by the models.

The observed amplitude of the rate variation is in phase with the MCEq predictions with stronger deviations in January, February, April, and September. The rate variation amplitudes calculated with the empirical model NRLMSISE-00 are up to 1\% larger in January and February, and about 1\% lower in September and October compared to the data-driven atmospheric parameterizations. 
A $\chi^2$-test incorporating only statistical uncertainties reveals a deviation of the observed rate from the calculation based on the NRLMSISE-00 model at $4.6\sigma$. The parameterizations from AIRS and ECMWF describe the data slightly better, particularly during the Austral summer months and in October, with p-values corresponding to $3.1\sigma$ for AIRS and $2.7\sigma$ for ECMWF.

These findings are underlined by the analysis in \cite{IceCube:2023qem}, determined from data within  the peak summer and winter months only.
It is important to note that this comparison of observed rates with calculated predictions is based on monthly averages. Additionally, the calculation of rate variation from the atmospheric parameterizations does not account for measurement or model uncertainties. The expected variation in spring and fall could be attributed to the modulation of the complex cooling and heating phases.

For the subsequent energy-dependent analysis, Austral summer and winter are defined as the periods with comparable amplitudes: October to January and May to August, respectively.

\section{Spectrum unfolding}\label{sec:unfolding}

The true neutrino energy distribution $f(E)$ cannot be determined directly since neutrino energies are inferred from the measurement of Cherenkov photons produced by neutrino-induced muons. The relationship between the true energy distribution and the observed distribution
 $g(y)$ of measurable quantities $y$ is described by the 
Fredholm integral equation of the first kind \cite{fredholm03}:

\begin{equation}
    g(y) = \int_a^c A(E,y) f(E) \mathrm{d}E + b(y)+  \epsilon(y).
    \label{fredholmequation}
\end{equation}
$A(E,y)$  represents the detector response function that maps the true energy distribution. The additional terms $b(y)$ and  $\epsilon(y)$ account for background contamination in the event selection and statistical/systematic uncertainties, respectively. Equation~\ref{fredholmequation} is discretized due to the finite number of observations:

\begin{equation}
    \vec{g}(y) = \textbf{A}(E,y)\vec{f}(E) + \vec{b}(y)+ \vec{\epsilon}(y). \label{equ:discrete_fredholm}
\end{equation}
Converting from measured quantities to the energy distribution constitutes an inverse problem, inherently ill-conditioned. The energy deposited by muons in the ice or nearby bedrock during their propagation is non-deterministic. At energies below \SI{1}{\tera\electronvolt}, continuous losses dominate, while stochastic losses dominate at higher energies, contributing to the smearing of energy resolution. Calculating particle propagation and energy losses relies on the lepton propagation tool PROPOSAL \cite{Koehne:2013gpa}.
Spectrum unfolding addresses this ill-conditioned problem by estimating the true distribution from measured proxies obtained from a Monte Carlo detector response simulation.

\subsection{Estimation of the energy spectrum} 
In this paper, we apply the DSEA algorithm to determine the neutrino energy spectrum for each season. 
 DSEA treats each energy bin (components of $\vec{f}(E)$ as described in Eq.~\ref{equ:discrete_fredholm}) as a distinct class of events, effectively transforming the spectrum estimation problem into a multinomial classification task.

The algorithm is trained through an iterative process on two training sets  each weighted according to the MCEq prediction for the annual average.  Initially, a random forest classifier is trained on energy-related features, discussed in the next section.
To avoid overfitting and assess generalization performance, the trained classifier is applied to a second, independent training set, predicting confidence scores $c_{M}(i\mid x_{\mathrm{n}})$, which indicate the likelihood that event $x_{\mathrm{n}}$ has an energy corresponding to bin index $i$. 

In the first iteration, the resulting confidence scores are reweighted to follow a uniform energy distribution, preventing bias from the classifier’s initial class distribution. The estimate obtained from this iteration is scaled by a variable step size, which controls the convergence speed of the algorithm and the regularization of the resulting spectrum. In subsequent iterations, the previous estimate is used as a prior to reweight the confidence scores accordingly. This iterative procedure—applying the classifier to a second independent training set and scaling the resulting estimate—ensures that the initial uniform weighting acts only as a starting point and does not bias the final unfolded spectrum.

The training process terminates when the change in the estimated spectrum between successive iterations becomes negligible, as quantified by the Wasserstein distance. This convergence criterion typically results in termination after three to five iterations, beyond which further updates do not produce meaningful changes to the spectrum.

The estimated content of each energy bin $\hat{f}_i$ after $k$ iterations is then obtained by applying the trained model to the seasonal data sets:

\begin{equation}
\hat{f_i}^{(k)} = \frac{1}{N_{\mathrm{events}}} \sum_{n=1}^{N_{\mathrm{events}}} c_{M}(i\mid x_{\mathrm{n}}) \qquad \forall \: 1 \leq i \leq I.
\label{equ:dsea}
\end{equation}
The sum of the confidence scores is normalized to $N_{\mathrm{events}}$ events in the data set, and $I$ denotes the total number of energy bins.
Detailed information on the algorithm is provided in \cite{2019ASPC..521..394R}.

\subsection{Energy proxies}\label{sec:proxies}

\begin{figure}[tbh]
    \centering
    \includegraphics[width=0.4\textwidth]{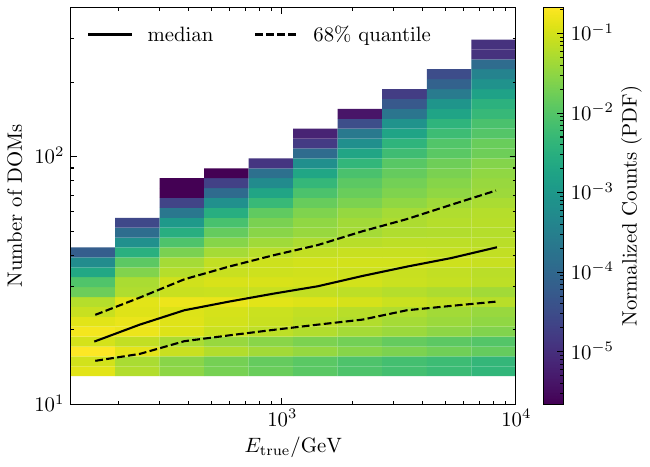}
    \includegraphics[width=0.4\textwidth]{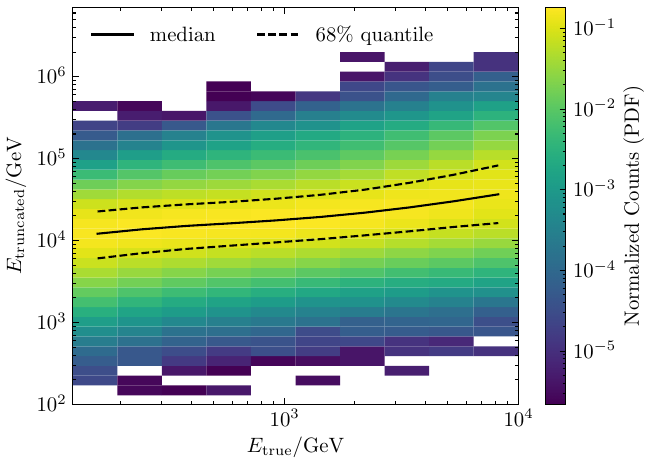}
    \caption{Correlation between energy proxies and simulated neutrino energy. The events are normalized to the total number of counts in the simulated dataset per column, binned according to true neutrino energy, as used in the analysis.}
    \label{fig:correlation}
\end{figure}

\begin{figure}
    \centering
    \includegraphics[width=0.4\textwidth]{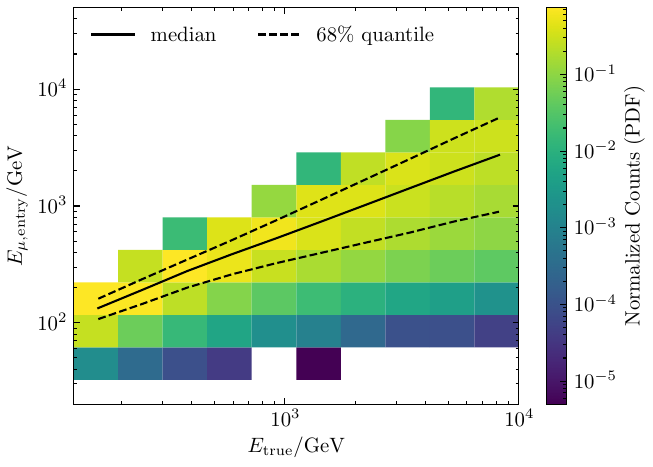}
    \caption{The correlation between the true muon energy at detector entry to the true neutrino energy. This smeared correlation is attributed to the unknown distance between the detector and the position of muon production, highlighting the complexity of accurately reconstructing neutrino energy from measured quantities.}
    \label{fig:emuon_eneutrino}
\end{figure}

The neutrino energy is estimated from two reconstructed proxies: the number of hit DOMs per event, referred to as \textit{number of DOMs}, and a likelihood reconstruction of the deposited energy along the muon trajectory through the detector, referred to as \textit{truncated energy} ($E_{\mathrm{truncated}}$)\cite{IceCube:2012iea}. The correlation between these proxies and the true neutrino energy is shown in Fig.~\ref{fig:correlation} for the energy range from \SI{125}{\giga\electronvolt} to \SI{10}{\tera\electronvolt} used in this analysis, including the median and the 68\% quantiles. The correlation of both quantities is smeared due to the unknown distance the muon traveled before entering the detector and the stochastic processes in muon propagation, as described above. This effect is highlighted in Fig.~\ref{fig:emuon_eneutrino}, which shows the correlation between the true neutrino energy and the true muon energy at the detector entry. Since the correlation is imperfect, the figure illustrates the complexity of estimating the neutrino energy from measurable detector quantities and highlights the necessity of an unfolding procedure to account for the smearing in the energy distribution.

\begin{figure}
\centering
\includegraphics[width=0.4\textwidth]{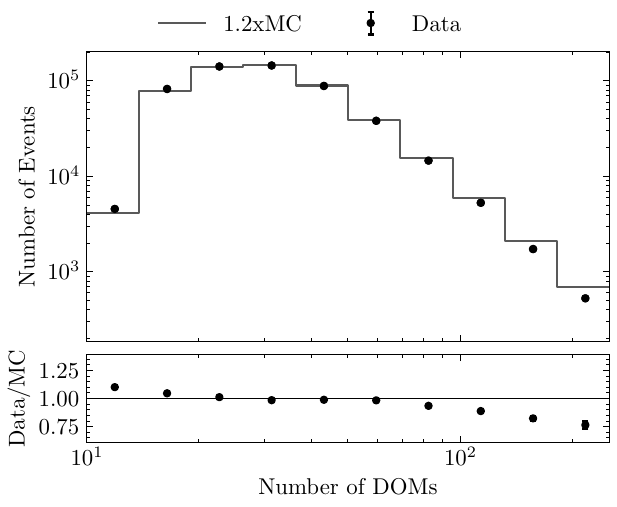}
\includegraphics[width=0.4\textwidth]{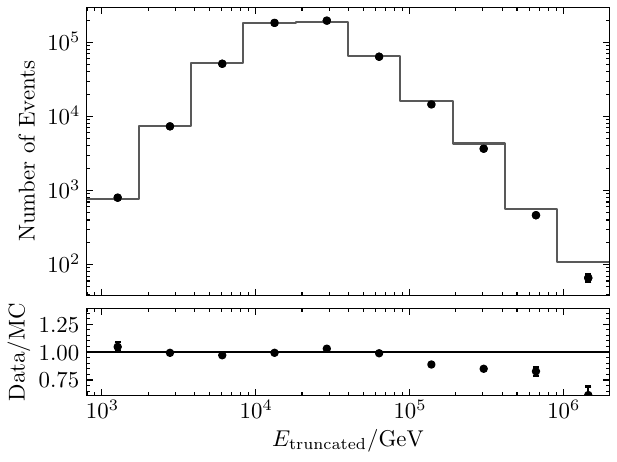}
\caption{Data-MC agreement for both energy proxies used in unfolding. The distribution of simulated events is weighted to MCEq and is normalized to the number of events in the data set by a normalization factor of 1.2. The top figure shows the proxy variable \textit{number of DOMs}, and the bottom figure shows the energy proxy \textit{truncated energy}.}
\label{fig:data-mc}
\end{figure}

Since the unfolding algorithm is trained on simulated events, a good agreement between the shape of simulated and experimental data is required for the energy proxies. The measurement of seasonal variations is robust against potential constant offsets in the distribution of simulated to experimental data since the measurement is performed relatively to the annual average flux, as further detailed in Section \ref{sec:results}. For algorithm training and data–simulation comparison, simulated events are weighted according to a realistic flux scenario for atmospheric and astrophysical muon neutrinos. The atmospheric flux assumption is provided by MCEq, with H3a as the primary composition, SIBYLL-2.3c as the hadronic interaction model, and NRLMSISE-00 as the atmospheric model for the annual average, as selected for the calculation shown in Fig.\ \ref{fig:mceq_prediction}. The astrophysical flux assumption is adopted from \cite{IceCube:2021uhz}. 

Fig.~\ref{fig:data-mc} shows the agreement between data and weighted simulation for both energy proxies. The distribution of simulated events is normalized to the total number of events in the dataset, resulting in a scaling factor of 1.2. The upper panel depicts the number of events per bin, including statistical uncertainties, while the lower panel shows the ratio between the two. The distribution of simulated events generally agrees with experimental data within statistical uncertainties, except for deviations observed in the number of DOMs above 100, and above $10^5\si{\giga\electronvolt}$ for $E_{\mathrm{truncated}}$. These respective ranges correspond to neutrino energies outside of the analyzed range below \SI{10}{\tera\electronvolt}, as can be inferred from Fig.~\ref{fig:correlation}. 

The algorithm is trained on the yearly average neutrino flux prediction and does not obtain any information on seasonal flux modulation. The robustness against deviations in the spectral index of the annual average to the training spectrum is tested, as described in \ref{pseudosample}. The energy spectrum is unfolded into ten logarithmically equidistant energy bins from \SI{125}{\giga\electronvolt} to \SI{10}{\tera\electronvolt}.

\subsection{Systematic and statistical uncertainties}\label{uncertainties}

\begin{figure*}
\centering
\includegraphics[width=0.9\textwidth]{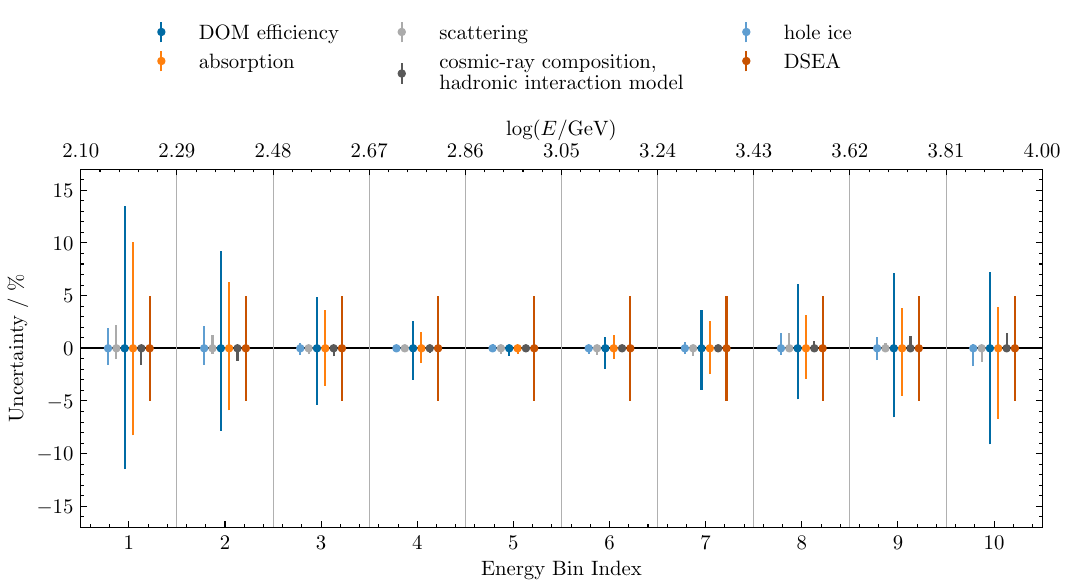}
\caption{A detailed overview of systematic uncertainties of the unfolded spectrum is provided per uncertainty type and energy bin. The systematic uncertainty remains constant across all seasons, as it is determined relatively to the reference simulation. The asymmetry of the uncertainties arises from the unfolding of pseudo data generated from simulations with altered systematic parameters, which have a larger effect on energy bins with smaller statistics.}
\label{fig:systematics}
\end{figure*}

Systematic uncertainties in the unfolded spectrum arise from various sources, including the detector uncertainties, detector medium, and reconstruction process. Each systematic uncertainty is assessed through simulated events, whereby pseudo-samples are unfolded by varying one parameter at a time.  This approach is consistent with previous uncertainty estimations in spectrum unfolding (e.g.,  \cite{IceCube:2014slq,IceCube:2017cyo,ANTARES:2021cwc}).

Specific uncertainties include those from quantum efficiency of the optical modules ($\sigma_{\mathrm{DOM}}$), ice absorption ($\sigma_{\mathrm{abs}}$), scattering ($\sigma_{\mathrm{scat}}$), optical properties of the refrozen ice around cable strings ($\sigma_{\text{hole ice}}$). These uncertainties are consistently calculated using variations of $\pm 10\%$ for DOM efficiency, $\pm 5\%$ for absorption and scattering, and polynomial function approximations for the hole ice \cite{Aartsen:2013rt,IceCube:2016zyt}. The flux uncertainties from hadronic interaction models and primary cosmic-ray composition ($\sigma_{\mathrm{flux}}$) are determined by parameterizing the results from   \cite{Fedynitch:2012fs}, accounting for changes in the spectral shape. The uncertainty associated with deviations from the true distribution in unfolded pseudo-samples  ($\sigma_{\mathrm{DSEA+}}$) is accounted for by constraining it as a constant systematic uncertainty of $\pm 5$\% across all energy bins (see Fig.\ \ref{fig:pseudosample} in \ref{pseudosample}).
The total uncertainty per energy bin is computed relative to the reference simulation, with upper and lower uncertainties assessed separately, summing all deviations in quadrature.

Statistical uncertainties are determined in a bootstrap approach, in which the dataset is sampled in 2000 trials and unfolded. Each sample is generated by randomly drawing events from the original dataset, allowing the same event to be selected more than once within a single sample. This procedure creates 2000 pseudo-samples that reflect the statistical distribution of the original dataset. Uncertainties per energy bin are derived from the covariance matrix, with statistical errors given by the square roots of its diagonal components.

Fig.~\ref{fig:systematics} shows the impact of each systematic uncertainty on the unfolded spectrum per energy bin, with larger uncertainties at the edges of the samples due to limited statistics. Asymmetric uncertainties can arise because the variation of a single parameter in the simulation does not necessarily yield a linear response in the unfolded spectrum. These systematic uncertainties are approximately seasonally independent as they arise from the in-ice detector, reconstruction and the data analysis pipeline.
The total and statistical uncertainties are listed in Table \ref{tab:flux} of the \ref{app:averageflux}.

The systematic uncertainties discussed in this section apply exclusively to the unfolded energy spectrum. A separate, energy-dependent bias —relevant only to the seasonal-to-annual flux ratio— is described in \ref{pseudosample}.

\section{Seasonal variations in the unfolded spectrum}\label{sec:results}

\begin{figure*}
    \centering
    \includegraphics[width=\textwidth]{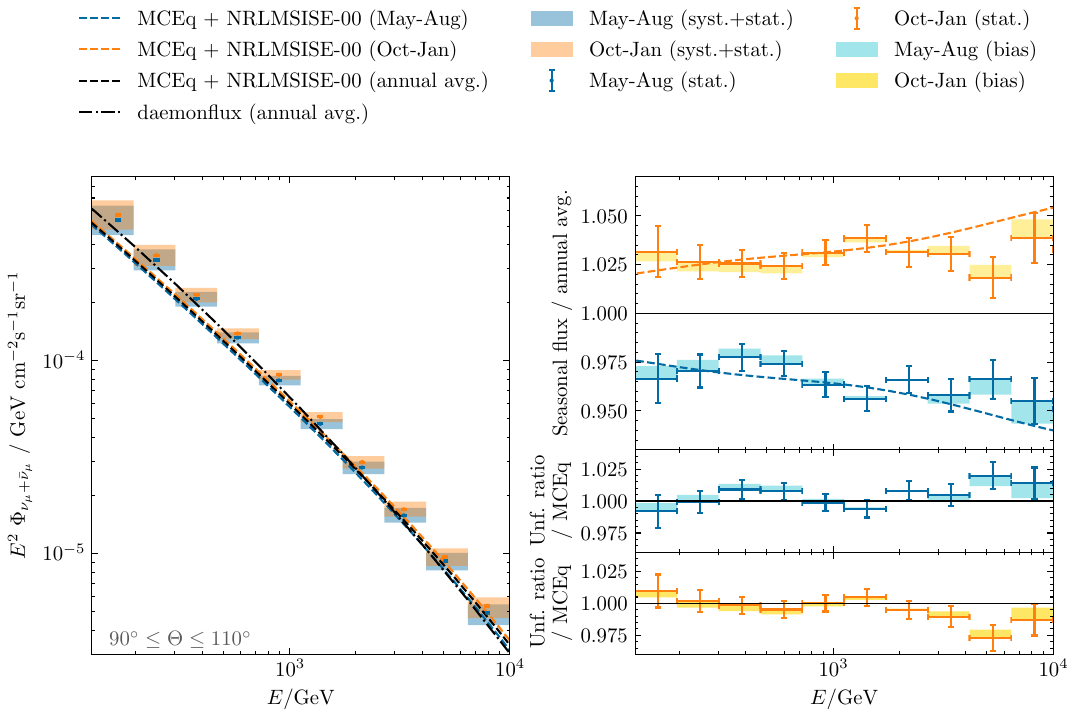}
    \caption{The left panel shows the unfolded seasonal fluxes for Austral summer (October to January) and Austral winter (May to August) in the zenith range from \SIrange{90}{110}{\degree}. Statistical uncertainties are depicted as error bars, while systematic uncertainties are represented as shaded areas. The theoretical predictions from MCEq (H3a, SIBYLL-2.3c, NRLMSISE-00) and daemonflux are shown in dashed and dash-dotted lines, respectively. The upper half on the right presents the ratio of the unfolded seasonal datasets to the annual average, indicating the seasonal variation strength across the energy range of interest. The systematic uncertainties remain constant for each season and cancel in the calculated ratio, leaving only statistical uncertainties. Additionally, a small energy-dependent bias in the seasonal-to-annual flux ratio is shown. The lower half on the right displays the deviation of the unfolded ratio of the seasonal fluxes to the annual average with respect to the MCEq predictions for the ratio.
}
    \label{fig:results}
\end{figure*}

The unfolded event spectra are converted to a differential flux by accounting for the livetime of the seasonal datasets, the effective area of the event selection determined from the detector response simulation, and the solid angle.
Fig.~\ref{fig:results} depicts the unfolded energy spectra corresponding to Austral summer (October to January) and winter (May to August) for the zenith range from \SIrange{90}{110}{\degree}. The left panel shows the unfolded seasonal spectra along with the corresponding statistical and systematic uncertainties, compared to MCEq with H3a as the primary composition, SIBYLL-2.3c as the hadronic interaction model, and NRLMSISE-00 as the atmospheric model, as well as to daemonflux. 

The unfolded seasonal spectra agree in shape with MCEq and daemonflux predictions within systematic uncertainties. Although, a detailed comparison of the unfolded annual average to both models reveals an excess in the data, as further detailed in Fig.\ \ref{fig:data_vs_models} in \ref{app:averageflux}, the ratios remain unaffected by normalization offsets, as all seasons exhibit a consistent excess relative to the models.

The upper half on the right in Fig.~\ref{fig:results} shows the ratio of unfolded seasonal fluxes to the annual average. As discussed in Section \ref{uncertainties}, systematic uncertainties cancel out in this relative measurement, leaving only statistical uncertainties in the unfolded seasonal-to-annual average flux ratio. A small energy-dependent bias, introduced by training the unfolding on the annual average and subsequently dividing the seasonally unfolded fluxes by the unfolded annual average, is also depicted in the figure and quantified in \ref{pseudosample}. The lower half on the right shows how the ratio of the unfolded seasonal fluxes to the annual average compares to MCEq predictions.

As expected, the amplitude of deviation of the seasonal to annual average flux increases with energy despite upward fluctuations in the first energy bin. During Austral summer, the rate variation rises from $(+3.2\pm 1.3)$\% at \SI{125}{\giga\electronvolt} to $(+3.9\pm 1.3)$\% at \SI{10}{\tera\electronvolt}. In contrast, during Austral winter, these ratios decrease from $(-3.3\pm 1.2)$\% to $(-4.5\pm 1.2)$\%. Table \ref{tab:variationamplitude} in the \ref{app:seasonalvariations} lists the seasonal variation amplitude per energy bin.

The bottom half on the right in Fig.~\ref{fig:results} depicts the ratio of the unfolded ratio of the seasonal to annual average flux to the  MCEq prediction with the NRLMSISE-00 atmospheric model separately for both seasons. The agreement is evaluated through a $\chi^2$-test, with p-values of 40.1\% for May-August and 26.4\% for October-January, indicating  consistency of the observed seasonal variation strength with model predictions compatible with the $1\sigma$ region according to the reduced $\chi^2$-values of 1.05 and 1.23 respectively.  

The choice of the hadronic interaction model has an impact on the spectral shape, but only a negligible impact on the seasonal variations below \SI{10}{\tera\electronvolt}, which is purely driven by the atmospheric parameterization.

\section{Conclusion and prospects}

This study presents an energy-dependent measurement of seasonal variations in the atmospheric muon neutrino flux within the zenith range from \SIrange{90}{110}{\degree}, covering energies from \SI{125}{\giga\electronvolt} to \SI{10}{\tera\electronvolt}. A previous measurement of the seasonal variation amplitude of atmospheric neutrinos from 6 years of IceCube data within the zenith range $\SI{90}{\degree} \leq \theta \leq \SI{115}{\degree}$ reported an amplitude of $\pm 3.5(3)$\%, which is in tension with model predictions of $\pm 4.3$\% using the AIRS atmospheric parametrization \cite{IceCube:2023qem}. Building upon this prior work, our findings indicate a seasonal variation amplitude of $(-3.5 \pm 0.2)$\% during the Austral winter (May to August) and $(+3.0 \pm 0.2)$\% during the Austral summer (October to January), averaged over the analyzed energy range. These results exhibit a comparable tension of $2.7\sigma$ with ECMWF temperature data, increasing to $3.1\sigma$ and $4.6\sigma$ for AIRS data and the atmospheric model NRLMSISE-00, respectively.

The energy-dependent analysis shows that the strength of seasonal variation increases with energy from October to January, reaching $(+3.9 \pm 1.3)$\%, and decreases with energy from May to August to $(-4.5 \pm 1.2)$\% relative to the annual average. This energy dependence stems from changes in atmospheric density gradients that influence production altitudes and the shift from pion-to kaon-dominated production. The unfolded seasonal variations in the energy spectrum align well with MCEq model predictions, using H3a as the cosmic-ray composition model, SIBYLL-2.3c for hadronic interactions, and the model NRLMSISE-00 for atmospheric conditions. While tensions are observed in the comparison of the observed rate variation with the MCEq prediction, no significant tension was found in the energy-dependent analysis for Austral summer and winter data. 
Discrepancies in rate variations relative to model predictions may arise from rapid temperature transitions during spring and fall, consistent with the findings in \cite{IceCube:2023qem}. This study highlights neutrinos as a unique probe for exploring variations in the Antarctic atmosphere  and the reliability of flux calculations with atmospheric models to describe the summer and winter seasons.

While the unfolded flux normalization within \SIrange{90}{110}{\degree} is up to 30\% higher than MCEq and daemonflux predictions, the differential seasonal variation measurements remain robust due to the cancellation of normalization uncertainties. This relative measurement allows to investigate the atmospheric neutrino flux in a regime often limited by significant modeling and measurement uncertainties.

Due to statistical limitations of the present dataset, the construction of seasonal datasets is based on averaging months with similar rate variations. For the same reason, seasonal unfolding above \SI{10}{\tera\electronvolt} is currently not feasible, preventing conclusions about a prompt component in the atmospheric neutrino flux or astrophysical neutrinos at comparable energies. Both are unaffected by seasonal variations and would reduce the observed amplitude for energies above the present range. Future analyses may investigate energy-dependent variations using a combination of atmospheric muons and muon neutrinos to measure a prompt component. The currently ongoing IceCube Upgrade \cite{Ishihara:2019aao} and IceCube-Gen2 expansion \cite{IceCube-Gen2:2020qha} will enhance model comparison capabilities by further reducing statistical and systematic uncertainties.

\begin{acknowledgements}

The IceCube Collaboration acknowledges the significant contributions to this manuscript by Karolin Hymon.
The authors gratefully acknowledge the support from the following agencies and institutions:
USA {\textendash} U.S. National Science Foundation-Office of Polar Programs,
U.S. National Science Foundation-Physics Division,
U.S. National Science Foundation-EPSCoR,
U.S. National Science Foundation-Office of Advanced Cyberinfrastructure,
Wisconsin Alumni Research Foundation,
Center for High Throughput Computing (CHTC) at the University of Wisconsin{\textendash}Madison,
Open Science Grid (OSG),
Partnership to Advance Throughput Computing (PATh),
Advanced Cyberinfrastructure Coordination Ecosystem: Services {\&} Support (ACCESS),
Frontera computing project at the Texas Advanced Computing Center,
U.S. Department of Energy-National Energy Research Scientific Computing Center,
Particle astrophysics research computing center at the University of Maryland,
Institute for Cyber-Enabled Research at Michigan State University,
Astroparticle physics computational facility at Marquette University,
NVIDIA Corporation,
and Google Cloud Platform;
Belgium {\textendash} Funds for Scientific Research (FRS-FNRS and FWO),
FWO Odysseus and Big Science programmes,
and Belgian Federal Science Policy Office (Belspo);
Germany {\textendash} Bundesministerium f{\"u}r Bildung und Forschung (BMBF),
Deutsche Forschungsgemeinschaft (DFG),
Helmholtz Alliance for Astroparticle Physics (HAP),
Initiative and Networking Fund of the Helmholtz Association,
Deutsches Elektronen Synchrotron (DESY),
and High Performance Computing cluster of the RWTH Aachen;
Sweden {\textendash} Swedish Research Council,
Swedish Polar Research Secretariat,
Swedish National Infrastructure for Computing (SNIC),
and Knut and Alice Wallenberg Foundation;
European Union {\textendash} EGI Advanced Computing for research;
Australia {\textendash} Australian Research Council;
Canada {\textendash} Natural Sciences and Engineering Research Council of Canada,
Calcul Qu{\'e}bec, Compute Ontario, Canada Foundation for Innovation, WestGrid, and Digital Research Alliance of Canada;
Denmark {\textendash} Villum Fonden, Carlsberg Foundation, and European Commission;
New Zealand {\textendash} Marsden Fund;
Japan {\textendash} Japan Society for Promotion of Science (JSPS)
and Institute for Global Prominent Research (IGPR) of Chiba University;
Korea {\textendash} National Research Foundation of Korea (NRF);
Switzerland {\textendash} Swiss National Science Foundation (SNSF).
\end{acknowledgements}
\printbibliography
\clearpage

\onecolumn
\begin{multicols}{2} 
\appendix

\section{Supplementary Material}\label{appendixa}
\subsection{Unfolded seasonal variation amplitude} \label{app:seasonalvariations}

\begin{table*}
\centering
\caption{Unfolded seasonal variation amplitude for Austral winter (May to August) and Austral summer (October to January) across the energy range.}
\label{tab:variationamplitude}       
\begin{tabular}{rrrrrrrrrrr}
\hline\noalign{\smallskip}
& & & & \multicolumn{4}{c}{Energy Bin Index} & \\
Season & 1 & 2 & 3 & 4 & 5 & 6 & 7 & 8 & 9 & 10 \\
\noalign{\smallskip}\hline\noalign{\smallskip}
May-Aug & $-3.3$\% &  $-3.0$\% &  $-2.2$\% &  $-2.6$\% &  $-3.7$\% &  $-4.4$\% &  $-3.4$\% & $-4.2$\% & $-3.4$\% &  $-4.5$\%  \\
$\pm$ (stat.) & 1.2\% & 0.8\% & 0.7\% & 0.6\% & 0.6\% & 0.6\% & 0.7\% & 0.8\% & 1.0\% & 1.2\% \\
 (bias) & $-0.6$\%  & $-0.6$\%  & $-0.4$\%  & $-0.4$\%  & $-0.3$\%  & $-0.1$\%  & $0.0$\%  & $+0.4$\% &$+0.8$\% & $+1.2$\% \\

\noalign{\smallskip}\hline\noalign{\smallskip}
Oct-Jan & $+3.2$\% & $+2.6$\% & $+2.6$\% & $+2.5$\% & $+3.1$\% & $+3.8$\% & $+3.1$\% & $+3.1$\%& $+1.8$\% & $+3.9$\%  \\
$\pm$ (stat.)& 1.3\% & 0.9\% & 0.7\% & 0.7\% & 0.7\% & 0.7\% & 0.8\% & 0.9\% & 1.0\% & 1.3\% \\
(bias) & $+0.5$\%  & $+0.5$\%  & $+0.4$\%  & $+0.3$\%  & $+0.2$\%  & $+0.2$\%  & $-0.1$\%  & $-0.4$\% &$-0.7$\% & $-1.0$\% \\
\noalign{\smallskip}\hline
\end{tabular}
\end{table*}

The unfolded seasonal variation amplitudes per energy bin for Austral summer and winter, corresponding statistical uncertainties and bias are listed in Table \ref{tab:variationamplitude}.

\subsection{Unfolded annual average muon neutrino flux}\label{app:averageflux}
\begin{figure*}
    \centering
    \includegraphics[width=\textwidth]{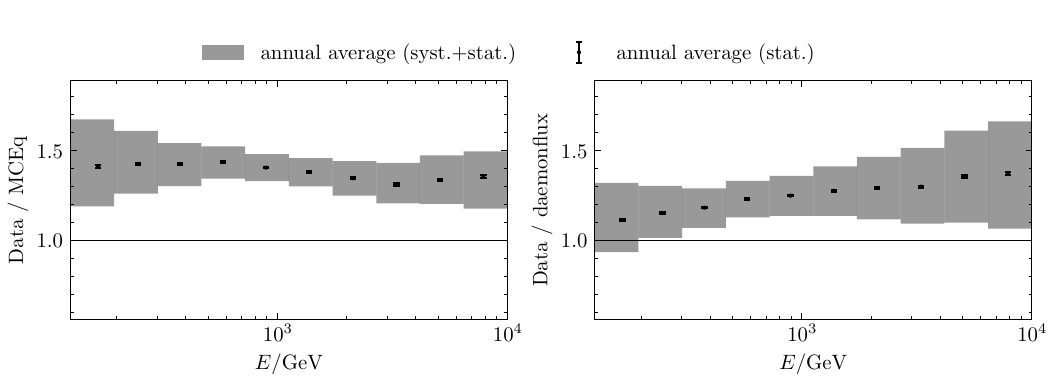}
    \caption{Ratio of the unfolded annual average flux to model predictions. The left panel displays the comparison to MCEq with H3a, SIBYLL-2.3c, while the right panel compares it to daemonflux. The shaded areas depict the statistical and systematic uncertainties of the unfolded flux. Additional uncertainties for the daemonflux model are included in the right panel.}
    \label{fig:data_vs_models}
\end{figure*}

The unfolded annual average flux used for the calculation of the seasonal variation strength in Fig.~\ref{fig:results} and corresponding statistical and systematic uncertainties are summarized in Table \ref{tab:flux}. The comparison of the unfolded annual average flux to MCEq and daemonflux from Fig. \ref{fig:results} is displayed in Fig.~\ref{fig:data_vs_models}. The shaded areas depict statistical and systematic uncertainties. The left panel showing the comparison to daemonflux takes into account additional systematic uncertainties from the data-driven model derivation.

\begin{table*}
\centering
\caption{Unfolded annual average atmospheric muon neutrino flux in the zenith range from \SIrange{90}{110}{\degree}, along with the corresponding statistical and systematic uncertainties.}
\label{tab:flux}       
\begin{tabular}{clcccll}
\noalign{\smallskip}\hline\noalign{\smallskip}
Energy Bin & $\log (E/\si{\giga\electronvolt})$  & Center energy & $ E^2 \frac{\mathrm{d}\phi}{\mathrm{d}E}$   & Stat. & \multicolumn{2}{c}{Syst. Uncertainty}\\
Index  & &   & $\left[\si{\giga\electronvolt\per\centi\metre\squared\per\second\per\steradian}\right]$& Uncertainty & &\\ 
\noalign{\smallskip}\hline\noalign{\smallskip}
1 & $2.10-2.29$ & $2.211$ & \num{5.55e-4} & $\pm0.6$\% & $-15.2$\% & $+17.8$\% \\
2 & $2.29-2.48$ & $2.397$ & \num{3.45e-4} & $\pm0.4$\% & $-11.2$\%  & $+12.5$\%\\
3 & $2.48-2.67$ & $2.584$ & \num{2.15e-4} & $\pm0.3$\% & $-8.2$\% & $+7.8$\% \\
4 & $2.67-2.86$ & $2.771$ & \num{1.36e-4} & $\pm0.3$\%  & $-6.0$\% & $+5.8$\%\\
5 & $2.86-3.05$ & $2.958$ & \num{8.22e-5} & $\pm0.3$\% & $-5.1$\% & $+5.0$\% \\
6 & $3.05-3.24$ & $3.146$ & \num{4.94e-5} & $\pm0.3$\% & $-5.5$\% &  $+5.3$\% \\
7 & $3.24-3.43$ & $3.335$ & \num{2.90e-5} & $\pm0.3$\% & $-6.9$\% & $+6.7$\% \\
8 & $3.43-3.62$ & $3.543$ & \num{1.65e-5} & $\pm0.3$\% & $-7.6$\% & $+8.8$\% \\
9 & $3.62-3.81$ & $3.713$ & \num{9.46e-6} & $\pm0.4$\% &  $-9.5$\% & $+10.0$\%\\
10 & $3.81-4.00$ & $3.903$ &  \num{5.16e-6} & $\pm0.6$\% & $-12.6$\% & $+10.0$\% \\
\noalign{\smallskip}\hline
\end{tabular}
\end{table*}

\subsection{Data-MC-agreement with systematic uncertainties}

\begin{figure*}
\centering
\includegraphics[width=0.48\textwidth]{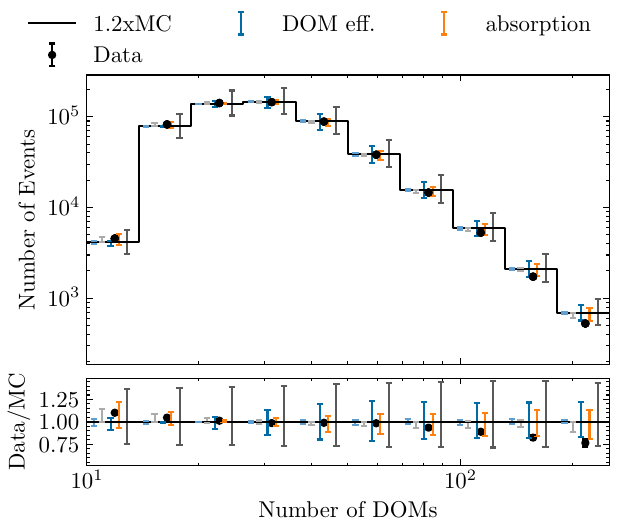}
\includegraphics[width=0.48\textwidth]{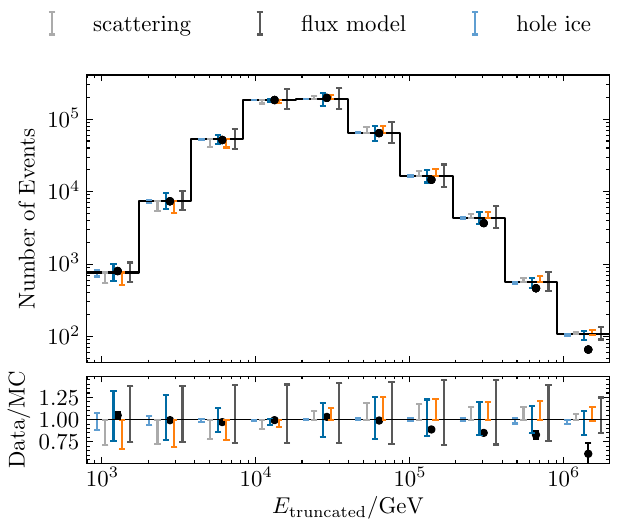}
\caption{Data-MC agreement with systematic uncertainties for both energy proxies used in unfolding. Error bars reflect the impact of varying individual systematic parameters relative to the baseline simulation. The distribution of simulated events is normalized to the number of events in the data set by a normalization factor of 1.2. }
\label{fig:data-mc-syst}
\end{figure*}

To provide a detailed assessment of the agreement between data and simulation across the unfolding proxies, the data-MC-comparison including systematic uncertainties is shown in Fig. \ref{fig:data-mc-syst}.  The error bars on the simulated distributions represent deviations from the baseline simulation arising from variations in individual systematic parameters, as detailed in Section~\ref{uncertainties}. The level of agreement between data and simulation is consistent with the estimated systematic uncertainties.

\section{Testing the unfolding method on pseudo-data}
\label{pseudosample}

As a methodological test, in addition to an annual average sample, two seasonal pseudo-samples—representing summer and winter—are generated by sampling events from the respective MCEq-predicted seasonal fluxes. Each pseudo-sample contains approximately 1 million events to ensure that the resulting distributions closely follow the MCEq fluxes and are minimally affected by statistical fluctuations.

The training spectrum consists of 1 million simulated neutrino events in the zenith range from \SIrange{90}{120}{\degree}, whereas the pseudo-sample covers the zenith range from \SIrange{90}{110}{\degree} only, matching the zenith range of the unfolded data set in Fig.~\ref{fig:results}. As discussed in Ref.~\cite{IceCube:2023ezh}, the zenith range was restricted from \SIrange{90}{120}{\degree} to \SIrange{90}{110}{\degree} due to the small average seasonal variation amplitude (below $\pm1$\%) observed in the \SIrange{110}{120}{\degree} zenith band, based on atmospheric temperature profiles from AIRS and ECMWF.

\begin{figure*}
    \centering
    \includegraphics[width=\textwidth]{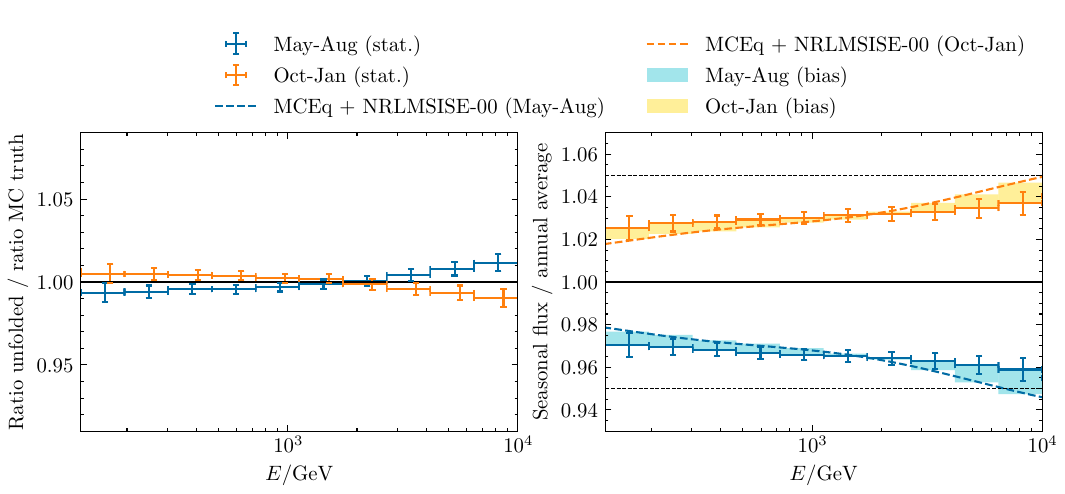} 
    \caption{The left panel shows the ratio of the 
    unfolded seasonal flux to the annual average, divided by the corresponding MC truth ratio obtained from the seasonal pseudo-samplescorresponding to \num{11.3} years of experimental data, with statistical uncertainties depicted as error bars. The mean of this ratio across pseudo-samples defines the unfolding bias, quantifying the systematic deviation from the MC truth. The right panel shows the ratio of the unfolded seasonal pseudo-samples to the annual average. Dashed lines for the respective seasons represent the theoretical predictions from MCEq (H3a, SIBYLL-2.3c, NRLMSISE-00). The bias from the left panel is depicted as shaded bands.}
    \label{fig:pseudosample}
\end{figure*}

The left panel of Fig.~\ref{fig:pseudosample} shows the ratio of the unfolded seasonal flux to the annual average, divided by the corresponding MC truth ratio obtained from the seasonal pseudo-samples. The statistical uncertainties of the unfolded pseudo-samples are calculated using a bootstrap approach detailed in Section \ref{uncertainties}. The unfolded ratio of the two seasonal pseudo-samples exhibits a flatter energy dependence compared to the MC truth, resulting in deviations of up to $\pm 1$\% in the three lowest and highest energy bins (listed in Table) \ref{tab:variationamplitude}. Notably, the slope of this deviation reverses in sign when switching between seasons, indicating a systematic trend not fully accounted for by statistical uncertainties. This suggests that the unfolding procedure introduces a small energy-dependent bias relative to the MC truth. The bias is added to the unfolded ratio of seasonal to annual average flux in Fig.~\ref{fig:results}.
As a conservative estimate, the uncertainty of the unfolded spectrum arising from the unfolding method is estimated by an upper bound of $\pm 5\%$, as depicted in Fig.~\ref{fig:systematics}.

The right panel of Fig.~\ref{fig:pseudosample} compares the unfolded seasonal-to-annual average flux ratio with the corresponding MCEq predictions, which are consistent with the MC truth. The bias given by the right panel is displayed in shaded bands, overlayed with statistical uncertainties from the unfolding. The unfolded seasonal variation strength, shown as the seasonal to annual average flux ratio, is consistent with the predicted ratios from MCEq for both seasons within statistical uncertainties and added bias. Adding the seasonal bias ensures that the unfolding method accurately recovers the seasonal variation signal in the restricted zenith range, without requiring retraining of the algorithm.

\end{multicols} 

\end{document}